\documentclass[aps,prd,twocolumn,superscriptaddress,groupedaddress]{revtex4} 

\usepackage{graphicx}  
\usepackage{dcolumn}   
\usepackage{bm}        
\usepackage{amssymb}   
\usepackage{epsfig}
\usepackage{epstopdf} 
\usepackage{subfigure}
\usepackage{mdframed}
\usepackage{amsmath}
\usepackage[colorlinks=true]{hyperref}  
\hypersetup{
    bookmarks=true,         
    unicode=false,          
    pdftoolbar=true,        
    pdfmenubar=true,        
    pdffitwindow=false,     
    pdfstartview={FitH},    
    pdftitle={My title},    
    pdfauthor={Author},     
    pdfsubject={Subject},   
    pdfcreator={Creator},   
    pdfproducer={Producer}, 
    pdfkeywords={keyword1} {key2} {key3}, 
    pdfnewwindow=true,      
    colorlinks=true,       
    linkcolor=magenta, 
    citecolor=blue,        
    filecolor=magenta,      
    urlcolor=cyan           
} 

\newcommand{\labell}[1]{\label{#1}}

\newcommand{\bea}{\begin{eqnarray}}
\newcommand{\eea}{\end{eqnarray}}
\newcommand{\ba}{\begin{eqnarray}}
\newcommand{\ea}{\end{eqnarray}}

\newcommand{\beq}{\begin{equation}}
\newcommand{\eeq}{\end{equation}}
\newcommand{\beqa}{\begin{eqnarray}}
\newcommand{\eeqa}{\end{eqnarray}}
\newcommand{\beqar}{\begin{eqnarray*}}
\newcommand{\eeqar}{\end{eqnarray*}}

\newcommand{\reef}[1]{(\ref{#1})}
\newcommand{\ssc}{\scriptscriptstyle}
\newcommand{\eg}{{\it e.g.,}\ }
\newcommand{\ie}{{\it i.e.,}\ }

\newcommand{\mt}[1]{\textrm{\tiny #1}}
\newcommand{\ang}[1]{\left\langle #1 \right\rangle}

\newcommand{\tL}{\tilde{L}}

\newcommand{\pd}{\partial}

\newcommand{\te}{t_\mt{E}}

\newcommand{\sg}{\sigma}
\newcommand{\q}{a}
\newcommand{\qe}{\q_{\rm\ssc E}}
\newcommand{\see}{S} 
\newcommand{\ka}{\kappa}
\newcommand{\kae}{\ka_{\rm \ssc E}}
\newcommand{\sge}{\sg_{\rm \ssc E}}

\newcommand{\ctt}{C_{\ssc T}}
\newcommand{\ctte}{C_{\ssc T\!, {\rm E}}}

\newcommand{\mc}{\mathcal}
\newcommand{\al}{\alpha}
\newcommand{\de}{\delta}
\newcommand{\ts}{\thinspace{}}
\newcommand{\req}[1]{(\ref{#1})} 

\begin{document}



\title{Universality of corner entanglement in {conformal field theories}} 
\author{Pablo Bueno}
\affiliation{Instituto de F\'isica Te\'orica UAM/CSIC, Nicol\'as Cabrera, 13-15, C.U. Cantoblanco, E-28049 Madrid, Spain}
\author{Robert C. Myers}
\affiliation{Perimeter Institute for Theoretical Physics, Waterloo, Ontario N2L 2Y5, Canada}
\author{William Witczak-Krempa}
\affiliation{Perimeter Institute for Theoretical Physics, Waterloo, Ontario N2L 2Y5, Canada}


\begin{abstract}
We study the contribution to the entanglement entropy of (2+1)-dimensional conformal field theories coming from a sharp corner in the entangling surface. This contribution is encoded in a function $a(\theta)$ of 
the corner opening angle, and was recently proposed as a measure of the degrees of freedom in the underlying CFT.
We show that the ratio $a(\theta)/\ctt$, where $\ctt$ is the central charge in 
the stress tensor 
correlator, is an almost universal quantity for a broad class of theories including various higher-curvature 
holographic models, free scalars and fermions, and Wilson-Fisher fixed points of the $O(N)$ models with $N=1,2,3$. 
Strikingly, the agreement between these different theories becomes exact in the limit $\theta\!\to\!\pi$, where the entangling 
surface approaches a smooth curve. We thus conjecture that the corresponding ratio is universal 
for general CFTs in three dimensions. 
\end{abstract}

\maketitle

Many interacting gapless quantum systems do not possess simple particle-like excitations, 
making it difficult to quantify their effective number of degrees of freedom (dof) at low-energy. 
Conformal field theories (CFTs) constitute an important example. For CFTs in two spacetime dimensions (2d), 
the Virasoro central charge is a good measure of the dof. It 
appears in many quantities, such as the thermal free energy and the entanglement entropy (EE), 
and decreases under renormalization group (RG) flow \cite{Zamolodchikov:1986gt}. In higher dimensions, the concept of quantum entanglement is emerging as a fundamental diagnostic for such   
measures \cite{Myers:2010xs,Myers:2010tj}. E.g.\/, it was instrumental in finding an analogous RG monotone for 
3d CFTs, with the EE of a disk-shaped region \cite{Casini:2012ei}. We shall study another measure 
of recent interest \cite{Casini:2008as,Casini:2009sr,Casini:2006hu,PhysRevLett.110.135702,pitch,2014arXiv1401.3504K,
Fradkin:2006mb,Hirata:2006jx,Guifre2009,Singh2012,Helmes2014,Helmes2014_2}: the coefficient capturing the contribution of sharp corners to spatial entanglement.     

\begin{figure}[h]
    \includegraphics[width=0.41\textwidth]{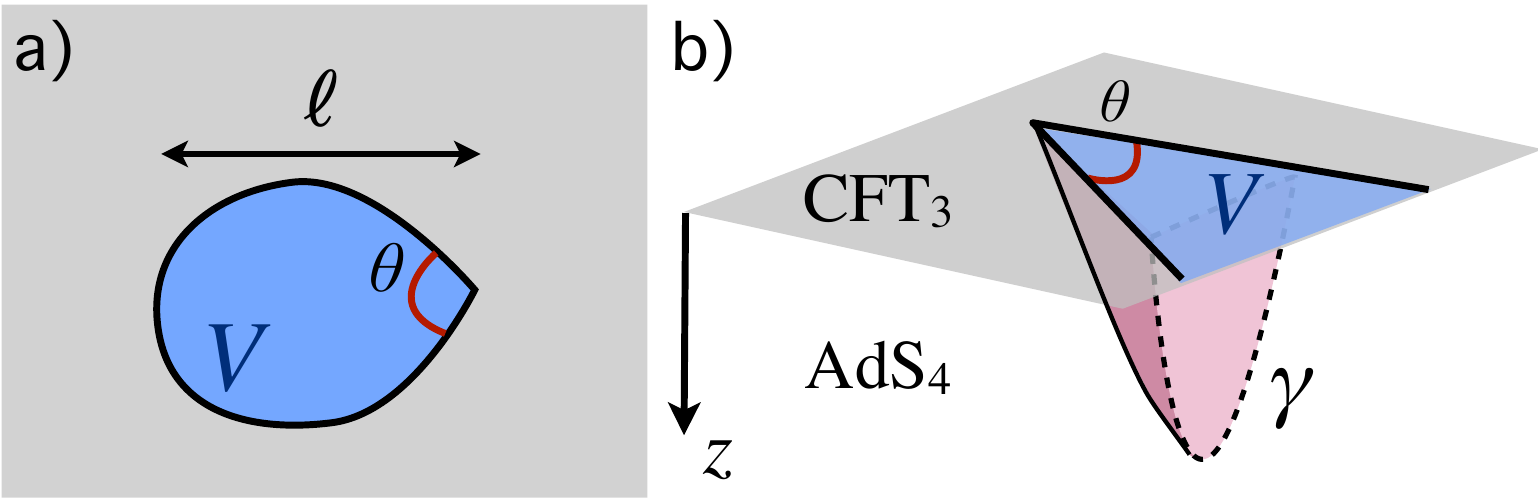}
    \caption{a) An entangling region $V$ of size $\ell$ with a corner; b) The holographic entangling surface $\gamma$ for a region
on the boundary of AdS$_4$ with a corner.}
    \label{fig0}
\end{figure}

In the context of quantum field theory, the EE is defined for a spatial region $V$ as: 
$S\!=\!-{\rm Tr}\left( \rho_V \ln \rho_V \right)$, where $\rho_V$ is the reduced density matrix produced by integrating out the dof in the complementary region $\overline{V}$. In the groundstate of a 3d CFT, the EE takes the form:  
\beq 
\see=B\,\ell/\delta-a(\theta)\,\ln\!\left(\ell/\delta \right)+\mathcal{O}(1) \, ,\labell{eek5}
\eeq
where $\delta$ is a short-distance cutoff, \eg the lattice spacing, and $\ell$, a length scale associated with the size of $V$. The first, `area law', term
depends on the UV regulator and scales with the size of the boundary. The second one appears only when $V$ has a
sharp corner with opening angle $\theta\!\in\![0,2\pi)$, Fig.\ts\ref{fig0}. 
Crucially, $\q(\theta)$ is a regulator independent coefficient that characterizes the underlying CFT. 
It is positive and satisfies $\q(2\pi-\theta)\!=\q(\theta)$ \cite{Casini:2008as}, 
and behaves as follows:
\beq \labell{sigma}
\q(\theta\rightarrow \pi)\simeq\sigma\ (\pi-\theta)^2\;\; ;\qquad 
 \q(\theta\rightarrow 0)\simeq \kappa/\theta\,
\eeq
in the limits of a nearly smooth entangling surface and a very sharp corner, {respectively}. {It} has been studied for a variety of systems: free scalars and fermions \cite{Casini:2008as,Casini:2006hu,Casini:2009sr},  interacting scalar theories via numerical simulations \cite{PhysRevLett.110.135702,pitch,2014arXiv1401.3504K}, Lifshitz quantum critical points \cite{Fradkin:2006mb}, 
and holographic models \cite{Hirata:2006jx}. The results suggest that $\q(\theta)$ is an effective measure of the dof 
in the underlying CFT \cite{Casini:2006hu,2014arXiv1401.3504K}.

Another quantity measuring dof is the central charge $\ctt$, associated with 
the stress tensor $T_{\mu\nu}$ of the CFT. 
It characterizes the vacuum two-point function:
\begin{align}\label{CT} 
 \langle T_{\mu\nu}(x)\, T_{\lambda\rho}(0) \rangle = \frac{\ctt}{|x|^{2d}}\, \mathcal I_{\mu\nu,\lambda\rho}(x)\,, 
\end{align}
where $\mathcal I_{\mu\nu,\lambda\rho}$ is a dimensionless tensor structure fixed by symmetry \cite{Osborn:1993cr}.
In the following, we will show that the ratio $\q(\theta)/\ctt$ is almost universal for a broad class of theories. In fact, we find that this agreement becomes exact in the limit $\theta\rightarrow \pi$. Hence, using \req{sigma} we conjecture there is a universal ratio, \ie the same value arises in any general 3d CFT: 
\begin{align}\labell{unisigct}
 \sigma/\ctt = \pi^2/24\, .
\end{align}
This is a striking result since the EE can generally be regarded as a nonlocal quantity but our analysis indicates that the  regulator independent corner contribution to the EE is controlled by a local correlation function \reef{CT}.

\begin{figure}
  \centering
    \includegraphics[width=3.42in]{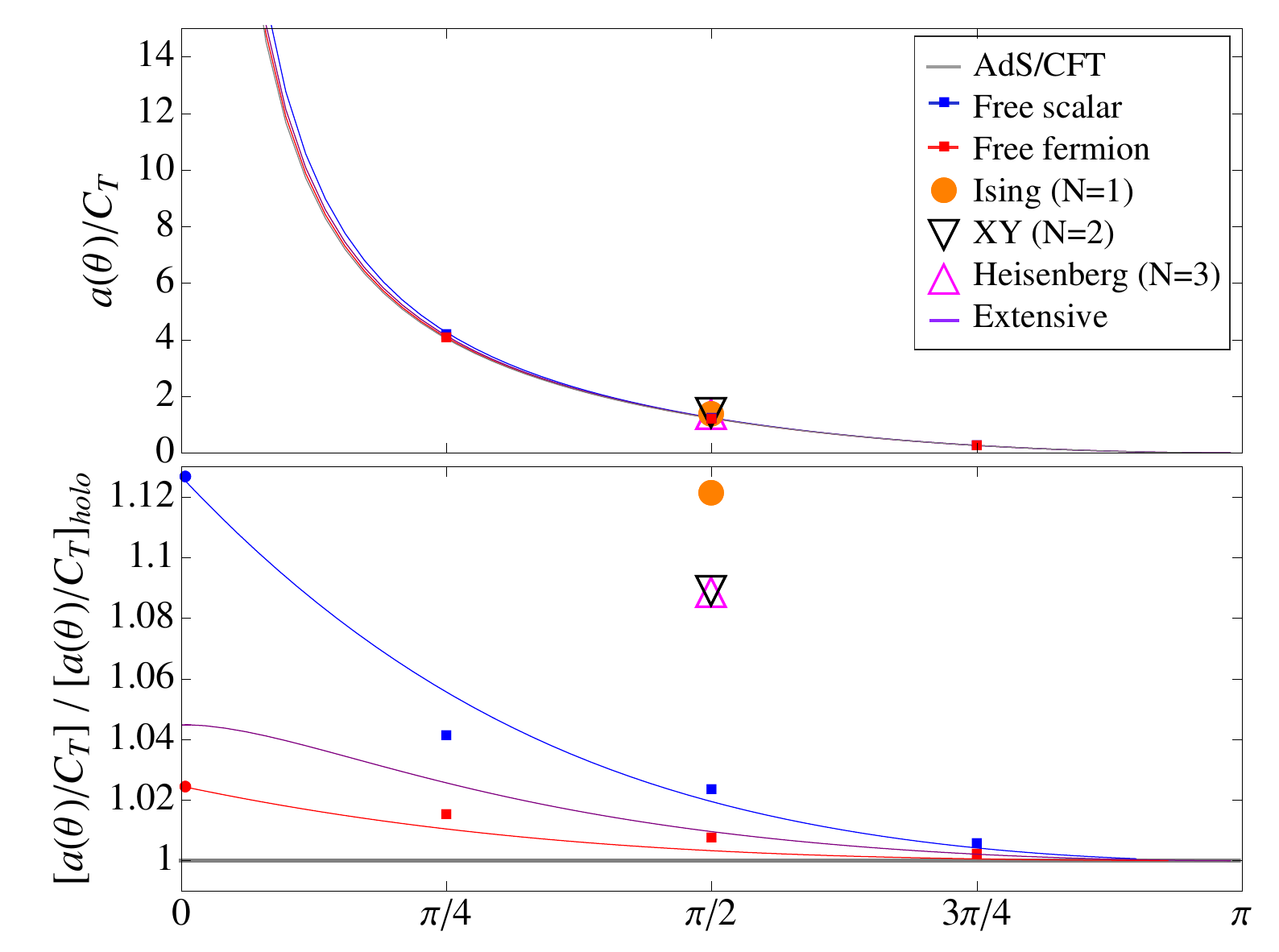} 
    \caption{(Top) $a(\theta)/\ctt$ from holography (gray), a free Dirac fermion (red) and a scalar (blue), plus the corresponding 
lattice data points obtained numerically \cite{Casini:2006hu} (red/blue squares). 
We also show $a(\pi/2)/\ctt$ for the $N\!=\!1,2,3$ Wilson-Fisher  
$O(N)$ CFTs, and the trial function \reef{Swingle} (purple).
(Bottom) Same quantities normalized by $[a(\theta)/\ctt]_{\rm holo}$.}  
    \label{fig1}
\end{figure}

\textbf{Holographic calculations:} The AdS/CFT correspondence posits that the physics of certain $d$-dimensional CFTs has an equivalent description in terms of gravity coupled to a negative cosmological constant in $d$+1 dimensions. In such holographic CFTs, EE is computed using the Ryu-Takayanagi prescription \cite{Ryu:2006bv,Ryu:2006ef}:
\beq
\see(V)=\underset{\gamma\sim V}{\text{ext}}\left[\frac
{\mathcal{A}(\gamma)}{4G} \right]\,. \labell{RyuTaka0} 
\eeq
That is, given a region $V$ in the boundary CFT, we consider all codimension-$2$ surfaces $\gamma$ in the dual AdS spacetime which are \emph{homologous} to $V$ on the asymptotic boundary and find the surface which extremizes the area functional within this class. The EE is then given by Bekenstein-Hawking formula $\mathcal{A}(\gamma)/(4G)$ evaluated on this extremal surface, where $G$ is the gravitational 
constant, Fig.\ts\ref{fig0}b. This prescription for the holographic EE (HEE) has been used to compute  the corner coefficient $\qe(\theta)$ for 3d CFTs dual to 4d AdS \cite{Hirata:2006jx,Myers:2012vs}; where `E' indicates that the bulk theory is described by Einstein gravity. 
$\qe(\theta)$ is only implicitly known in terms of 2 integrals (appendix \ref{holo})
but it is easily evaluated numerically and the result is plotted in Fig.\ts\ref{fig1}.  Evaluating $\qe(\theta)$ for the limits 
in \req{sigma}, we find:  $\kae=\Gamma({\tfrac34})^4\,\tL^2/(2\pi G)$ and  $\sge=\tilde{L}^2/(8\pi G)$, where $\tL$ is the radius of curvature of the dual AdS geometry \cite{BuenoMyers}.     
 
Just as for $\kae$ and $\sge$, $\qe(\theta)$ carries an overall factor of the ratio $\tL^2/G$ 
which is indicative of the number of dof in the boundary CFT. E.g.\/, \cite{Aharony:2008ug} provides a top-down model where the boundary CFT is a specific supersymmetric gauge theory and $\tL^2/G\sim N^{2}/\sqrt{\lambda}$ where $N$ and $\lambda$ are the rank 
of the gauge group and the 't Hooft coupling, respectively. 
However, this same ratio appears in many other physical quantities, \eg the stress tensor two-point function \reef{CT},     
EEs for general regions or the thermal entropy density. Essentially, this ubiquitous appearance of $\tL^2/G$ occurs because it is the only dimensionless parameter in the bulk Einstein theory. As a result, all of these a priori distinct measures of the dof are all related in this class of CFTs. However, by introducing 
higher curvature interactions, the bulk gravity acquires new dimensionless couplings and each of the above measures 
can acquire a distinct dependence on these couplings. Hence we can see whether the various measures are independent or if they encode the same information, \eg see \cite{JHxb,gb2,Myers:2010jv}.   

Higher curvature interactions appear generically in string theoretic models, \eg as $\alpha^{\prime}$ corrections in the low-energy effective action \cite{gross}. However, rather than constructing explicit top-down holographic models, our approach will be to examine simple toy holographic models involving higher curvature gravity in the bulk. 
Our perspective is that if there are interesting universal properties which hold for all CFTs, then they should also appear in the holographic CFTs defined by these toy models as well. This approach has been successfully applied before, \eg in the discovery of the F-theorem \cite{Myers:2010xs,Myers:2010tj}.

In particular, we focus our attention on the following simple gravitational theory
\begin{eqnarray}\labell{fo}
I= \int \frac{d^{4}x \, \sqrt{g}}{16\pi G}\left[\frac{6}{L^2}+R+ L^4\lambda_1 R \mathcal{X}_4+L^6\lambda_2 \mathcal{X}_4^2\right]\,.
\end{eqnarray}
The first two terms above are the cosmological constant with $\Lambda=-3/L^2$ and the standard Einstein term. The next two interactions are controlled by the dimensionless couplings $\lambda_{1,2}$ and contain $\mathcal{X}_{4}=R_{\mu\nu\rho\sigma}R^{\mu\nu\rho\sigma}-4R_{\mu\nu}R^{\mu\nu}+R^2$, which is the Euler density on 4d manifolds. While $\mathcal{X}_4$ alone would be topological, $R \mathcal{X}_4$ and $\mathcal{X}_4^2$ are not, and so these terms do modify the gravitational equations of motion. In particular, AdS space (with radius of curvature $\tL$) is a solution provided: $1=L^2/\tilde{L}^2-24\lambda_1L^6/\tilde{L}^6+96\lambda_2L^8/\tilde{L}^8$. Note that in the following, we treat the higher curvature interactions as perturbative corrections and we will only calculate to leading order in $\lambda_{1,2}$.

With higher curvature interactions, the Ryu-Takayanagi prescription \reef{RyuTaka0} must be 
modified \cite{highc2,Dong,Camps:2013zua,highcXX1,highcXX2}. Schematically, we have 
$\see(V)=\underset{\gamma\sim V}{\text{ext}}S_\mt{grav}(\gamma)$
where the new entropy functional contains the `higher curvature corrections' to the Bekenstein-Hawking formula. For the above action \reef{fo}, one finds \cite{Sarkar,BuenoMyers} 
\beq \labell{see3x} 
S_\mt{grav} \!=\!
\int_{\gamma }\!\! \frac{d^2y\, \sqrt{h}}{4G}\Big[1+\lambda_1 L^4 (\mathcal{X}_4+2R\mathcal{R})+4\lambda_2L^6\mathcal{X}_4\mathcal{R}\Big] \, ,
\eeq
where $h_{ij}$ and $\mathcal{R}$ are the induced metric and intrinsic Ricci scalar, respectively, on $\gamma$. 

{When} evaluating the HEE in the ground state of the boundary CFT, the bulk geometry is empty AdS$_{4}$, and \req{see3x} simplifies because $\mathcal{X}_4=24/\tL^4$ and $R=-12/\tL^2$ are constants. Further $\mathcal{R}$ is topological on two-dimensional manifolds, \ie $\sqrt{h}\mathcal{R}$ can be written as a total derivative. Thus, in this case, the HEE is still determined by an extremal area surface in the bulk and the corner term contribution is easily evaluated for the CFTs dual to \req{fo}:
\begin{align} \labell{alpha} 
 \q(\theta)=\alpha\, \qe(\theta)
  \quad{\rm with}\ \  \alpha= 1+24\lambda_1+\mathcal{O}(\lambda_i^2)\,.
\end{align}
Hence the corner coefficient is modified by an overall factor but the $\theta$-dependence is unchanged. 
The above result also implies $\kappa=\alpha\,\kae$ and $\sigma=\alpha\, \sge$.  

Now we compare this result to other measures of the dof in the holographic CFTs dual to the action \reef{fo}, beginning with the central charge $\ctt$. Since the stress tensor in the boundary CFT is dual to the metric perturbations in the bulk gravity theory, \req{CT} translates to a 
statement about the graviton propagator between two boundary points in AdS$_4$ \cite{hong22,gb2}. Hence we must determine the normalization of the graviton kinetic term in our higher curvature theory \reef{fo}. However, we should note that generically the higher curvature interactions introduce additional massive dof in the bulk theory with $M^2\sim 1/(\lambda_i L^2)$ \cite{Myers:2010tj,BuenoMyers}. Hence, the metric will couple to the stress tensor, but also to some additional (nonunitary) operators. However, in a perturbative framework, it is straightforward to determine $\ctt$ by computing the equation of motion corresponding to the massless spin-2 graviton mode. In particular, we consider metric
  fluctuations  $g_{\mu\nu}=\bar{g}_{\mu\nu}+h_{\mu\nu}$, where $\bar{g}_{\mu\nu}$ is the background AdS$_4$ metric 
and $h_{\mu\nu}\ll 1$. The equation of the physical massless graviton can be isolated by choosing the transverse gauge condition $\bar{\nabla}^{\mu}h_{\mu\nu}=\bar{\nabla}_{\nu}h$, 
supplemented by the tracelessness 
condition $\bar{g}^{\mu\nu}h_{\mu\nu}=0$, and reads
\beq \labell{eomg}
-\!\frac{\alpha}{2}\! \left[\bar{\Box}+\frac{2}{\tilde{L}^2}\right]\! h_{\mu\nu} =8\pi G \,{T}_{\mu\nu},
\eeq
where $\alpha$ is the same coefficient as in \req{alpha}. 
We have included the bulk stress tensor $T_{\mu\nu}$ of matter fields to establish the 
normalization of the graviton kinetic term. The net effect of the appearance of $\alpha$ in \req{eomg} is to modify the holographic 
result for the stress tensor correlator \reef{CT} by an overall factor $\alpha$:
\begin{align}
  \ctt=(1+24\lambda_1+\mathcal{O}(\lambda_i^2))\,\ctte \,,
\labell{house}
\end{align}
where the Einstein result is $\ctte=3\tilde{L}^2/(\pi^3G)$ \cite{hong22,gb2}.

Remarkably then, for all the holographic CFTs dual to \req{fo}, we find that $\q(\theta)/\ctt=\qe(\theta)/\ctte$, \ie we find a universal ratio that is independent of the details of the theory. It is notable that this universality does not occur when considering other measures of the boundary dof. Here we examined the coefficient in the thermal entropy density, $s=c_s\, T^2$, which can be determined by evaluating the horizon entropy of a planar AdS black hole, and the RG monotone $F$, which is determined using the HEE prescription with a circular entangling surface in the boundary. 
The results are
\begin{align}
c_s =& \left (1+24\lambda_1+16\lambda_2+\mathcal{O}(\lambda_i^2)\right)\,c_{s,\rm \ssc E}\,,
\nonumber\\
F =& \left(1+48\lambda_1-96\lambda_2+\mathcal{O}(\lambda_i^2) \right)\,F_{\rm\ssc E}\,,
\labell{house2}
\end{align}
where $c_{s,\rm\ssc E}=4\pi^2\tL^2/(9G)$ and $F_{\rm\ssc E}=\tL^2/(2G)$. Hence each of these measures of dof has a unique signature in terms of the couplings $\lambda_{1,2}$ and it is only the ratio $a(\theta)/\ctt$ that yields a universal result. Further in \cite{BuenoMyers}, this holographic analysis was extended to an 8-parameter family of higher curvature theories and the same universal ratio arises in the dual boundary CFTs.

\textbf{QFT comparison:}
The universality revealed by our holographic analysis suggests more broadly that $\q(\theta)/\ctt$ provides a useful normalization if we wish to compare the corner coefficients of different field theories. The only cases where $\q(\theta)$ is known for a wide range 
of angles are the free massless scalar and fermion \cite{Casini:2008as,Casini:2006hu,Casini:2009sr}. Despite being free field theories, the required calculations are technically demanding and $\q(\theta)$ is only given implicitly in terms of a complicated set of nonlinear differential and algebraic equations. Explicit results were given for Taylor expansions of $a(\theta)$ around $\theta=\pi$ (to 14th order), and for  
$\theta=\pi/4,\,\pi/2,\, 3\pi/4$ \cite{Casini:2006hu}, which were obtained using the numerical methods of \cite{Peschel}. The values of $\kappa$, \ie the coefficient at $\theta\to0$, can also be determined using a conformal transformation \cite{Casini:2009sr,BuenoMyers}. Finally analytic expressions for the free field central charges are known \cite{Osborn:1993cr}: $C_{\ssc T}^{\rm scalar}= 3/(32\pi^2)$ and $C_{\ssc T}^{\rm fermion}= 3/(16\pi^2)$. Combining these results, the ratios $\q(\theta)/\ctt$ for the holographic and free field theories are compared in Fig.\ts\ref{fig1}. The solid lines for the free fields are interpolating functions which combine the Taylor expansions around $\theta=\pi$ with the coefficients $\kappa$ at $\theta= 0$. 

Fig.\ts\ref{fig1} also includes $a(\pi/2)/\ctt$ for the Wilson-Fisher fixed points of the $O(N)$ models with $N=1,2,3$. In this case, $a(\pi/2)$ was evaluated numerically using state of the art numerical simulations for lattice Hamiltonians with the corresponding 
quantum critical points \cite{PhysRevLett.110.135702,pitch,2014arXiv1401.3504K}. Next, $\ctt$ was recently determined with great accuracy using conformal bootstrap methods \cite{Kos:2013tga}. 
Remarkably the lattice results indicated that $a(\pi/2)$ satisfies the emergent scaling $\q^{O(N)}(\pi/2)\simeq N\q^{\rm Ising}(\pi/2)$, to within the numerical accuracy. Further, the bootstrap calculations yield $\ctt$ with the same approximate scaling for $1\leq N\leq 3$. Hence
$a(\pi/2)/\ctt$ is nearly independent of $N$ for these CFTs, Table\ts\ref{tbl:ratio}.
Finally, Fig.\ts\ref{fig1} also shows a trial function, with a simple closed form, obtained from \cite{Casini:2008wt,Casini:2005rm,Swingle:2010jz}, see \req{Swingle}. 
\begin{table*}
  \centering
  \begin{tabular}{c||c|c|c|c|c|c} 
     &  Ising ($N\!=\!1$) & XY $(N\!=\!2)$  & Heisenberg ($N\!=\!3$) & Free scalar  & Dirac fermion & AdS/CFT\\
    \hline \hline
    $a(\pi/2)/\ctt$ &  1.36(14)\, \tiny{\cite{PhysRevLett.110.135702,Kos:2013tga}} & 1.3(1)\, \tiny{\cite{pitch,Kos:2013tga}} & 1.3(1)\, \tiny{\cite{2014arXiv1401.3504K,Kos:2013tga}} & 1.245\, \tiny{\cite{Casini:2006hu,Osborn:1993cr}} & 1.226\, \tiny{\cite{Casini:2008as,Osborn:1993cr}} & 1.222\, \tiny{\cite{Hirata:2006jx,hong22}} \\
    $a_2(\pi/2)/\ctt$ & 0.62(6)\, \tiny{\cite{Guifre2009,Humeniuk2012,Singh2012,Inglis2013,PhysRevLett.110.135702,Kos:2013tga} }& 0.62(6)\, \tiny{\cite{pitch,Helmes2014,Devakul2014,Kos:2013tga} }& 0.61(6)\, \tiny{\cite{Helmes2014_2,2014arXiv1401.3504K,Devakul2014,Kos:2013tga}} & 0.674\, \tiny{\cite{Casini:2006hu,Osborn:1993cr}}  & - & - 
  \end{tabular}
  \caption{Ratio $a_{n}(\pi/2)/\ctt$ for $n=1,2$ in different critical theories. The first 3 are the O$(N)$ Wilson-Fisher CFTs. 
}
\label{tbl:ratio}   
\end{table*}   
 
We see that all the results in Fig.\ts\ref{fig1} are in remarkable agreement 
for the whole range of $\theta$.  Refs.~\cite{Casini:2006hu,Nishioka:2009un} had previously noted a qualitative similarity between the corner coefficient in the holographic and free field theories. However, here we see a remarkable quantitative agreement after normalizing $\q(\theta)$ with the central charge $\ctt$. The free scalar curve differs from the holographic function by no more than $13\%$ in the whole range, while the agreement with the free fermion is even better, differing from the holographic result by no more than $2.5\%$. This excellent agreement also extends to the O$(N)$ Wilson-Fisher CFTs.
As we can see in Table \ref{tbl:ratio} and the lower plot in Fig.\ts\ref{fig1}, 
the ratio $a(\pi/2)/\ctt$ for these CFTs agrees remarkably well amongst each other and deviates only slightly from the holographic and free field results. 

A conspicuous feature shown in Fig.\ts\ref{fig1} is that the ratio $[a(\theta)/\ctt]\,/\,[a(\theta)/\ctt]_{\rm holo}$ is a monotonically decreasing function of the opening angle
for both free field theories. Hence the maximum discrepancy between these theories occurs at $\theta=0$. The most striking aspect of the plot is that these ratios approach 1 as   $\theta\!\to\!\pi$, indicating the possibility of universal behavior that extends beyond holography to general 3d CFTs.

\textbf{Universal ratio:}
As described in \req{sigma}, $a(\theta)$ is constrained to vanish quadratically at $\theta=\pi$, \ie in the limit of a smooth entangling surface.  Our holographic analysis indicates that the ratio of the corresponding coefficient with the central charge is  universal for a broad family of holographic CFTs (see also \cite{BuenoMyers}),
\beq
\sigma/\ctt=\sge/\ctte=\pi^2/24\simeq 0.411234\,.
\labell{cheese}
\eeq
Fig.\ts\ref{fig1} suggests that the free scalar and free fermion theories yield the same ratio and so we can evaluate this ratio for these theories, using the values of $\ctt$ given above and those of $\sigma$ given in \cite{Casini:2008as,Casini:2006hu,Casini:2009sr}: $\sigma_{\rm scalar}\simeq 0.0039063$, $\sigma_{\rm fermion}\simeq 0.0078125$. We find: $(\sigma/\ctt)_{\rm scalar}\simeq 0.411235$,  $(\sigma/\ctt)_{\rm fermion}\simeq 0.411234$.
Hence the free field ratios agree with the holographic result to at least five significant figures. Recall that the free field values of $\ctt$ are exact but the values of $\sigma$ are approximate numerical results \cite{Casini:2008as,Casini:2006hu,Casini:2009sr}. Therefore the precision of the agreement here is as good as possible. 

This motivated the conjecture \req{unisigct} that the ratio $\sigma/\ctt=\pi^2/24$ for all 3d CFTs. 
Our conjecture can be used to predict the \emph{exact} free field values for $\sigma$: 
\beq
\labell{kass}
\sigma_{\rm scalar}= 1/256\ ,\qquad
\sigma_{\rm fermion}= 1/128\ .
\eeq
To test this prediction, we revisited the original computations for these coefficients \cite{Casini:2008as,Casini:2006hu,Casini:2009sr}. In appendix \ref{cheese2}, we explicitly give the monstrous integrals needed to calculate $\sigma$ for the free fields. 
Improving the numerical accuracy in evaluating these integrals, we found that the agreement between the numerical results and \req{kass} was easily extended to 1 part in $10^{12}$ -- the accuracy to which we limited ourselves.\footnote{Remarkably enough, these integrals have been recently evaluated analytically \cite{Marios}, confirming our predictions in \req{kass}. } 
 
These two integrals 
are very complicated and not similar. 
The fact that they yield simple fractions and differ by a factor $2$ is highly non-trivial.

\textbf{Conclusions \& outlook:} We have shown that $a(\theta)/\ctt$ is an almost universal function of the opening angle for a broad class of CFTs, including a family of higher-curvature  holographic models, free massless scalars and fermions, and Wilson-Fisher fixed points of the $O(N)$ models with $N=1,2,3$ -- see Fig.\ts\ref{fig1}. A striking aspect of this result is that generally
the EE is considered a nonlocal quantity but here we found the regulator independent corner contribution to the EE encodes essentially the same counting of dof as the local correlation function \reef{CT}.\footnote{Note that  $\ctt$ is not an RG monotone in $d=3$ \cite{gum}.}

While the ratios $a(\theta)/\ctt$ for the different theories agree well across the whole range of $\theta$, the agreement becomes exact as $\theta\to\pi$. 
Given the extremely different nature of these CFTs and the computations 
involved in evaluating $a(\theta)$ in each case, we conjectured that
$\sigma/\ctt=\pi^2/24$ is a universal result for all 3d CFTs. Checking its validity in additional theories is an exciting issue to 
address. It seems possible to prove our conjecture \cite{prep} using the techniques developed in \cite{vlad1,vlad2}. Further, holographic calculations suggest that similar universal behavior may also arise in higher dimensions \cite{prep}.

Higher R\'enyi entropies also contain regulator independent corner functions $a_n(\theta)$, analogous to $a(\theta)$ (which corresponds to $n=1$). 
With results from lattice studies of the Wilson-Fisher CFTs \cite{PhysRevLett.110.135702,pitch,2014arXiv1401.3504K} and 
calculations for the free scalar field \cite{Casini:2006hu}, Table \ref{tbl:ratio} suggests that the almost universal 
behavior found for $a(\theta)/\ctt$ may extend to $a_n(\theta)/\ctt$ with $n\neq1$. Further, at $\theta=\pi$, the R\'enyi corner terms also have
a quadratic zero, $a_n(\theta\sim \pi )=\sigma_n (\pi-\theta)^2$. 
It would be interesting to determine if these coefficients are also universal, in the sense of \req{unisigct}. 
We have calculated the corresponding ratios in the free scalar theory  
for $n=2,3$: $\sigma_2/\ctt=2/9$ and $\sigma_3/\ctt=8\pi/(81\sqrt{3})$ (appendix \ref{cheese2}). 
It would be interesting to investigate these quantities in other theories, \eg for a free fermion.

While $a(\theta)$ is only known implicitly for the free fields \cite{Casini:2008as,Casini:2006hu,Casini:2009sr} and for holographic CFTs \cite{Hirata:2006jx}, the 
so-called \emph{extensive mutual information} model \cite{Casini:2008wt,Casini:2005rm,Swingle:2010jz} produces the simple closed form expression 
\begin{equation}\labell{Swingle}
 a_{\rm \ssc Ext}(\theta) = \pi^2 \ctt/8\ \left(1 + (\pi-\theta)\cot\theta\right)\,,
\end{equation}  
where we have fixed the overall normalization to recover \req{unisigct}. 
Interestingly, $\ctt$ arises from a pre-factor, which is the derivative of the 
scaling dimension of a twist operator with respect to the R\'enyi index. This derivative was recently shown to be $\pi^3\ctt/24$ for general CFTs \cite{Hung:2014npa} --
see appendix \ref{cheese3} for further comments.  This simple expression \req{Swingle} is also shown in Fig.\ts\ref{fig1} 
and exhibits the same nearly universal behavior as the other theories.

Finally, we observe that the holographic result for $\q(\theta)/\ctt$ takes the smallest values in Fig.\ts\ref{fig1} across the whole range of $\theta$.  The fact that the results for all the other CFTs considered, both interacting and free, persistently lie above the holographic curve is certainly remarkable. 
It would be of interest to investigate  whether the 
holographic result represents some kind of universal lower bound. This is reminiscent of the conjecture that $\eta/s$ is minimized by holographic
CFTs dual to Einstein gravity \cite{Kovtun:2004de}. In that case, however, the introduction of higher curvature terms was found to invalidate the claim \cite{Brigante:2007nu,steve2}, whereas $\sigma/\ctt$ remains unaffected by such terms. A challenge to this conjecture would arise from holographic models  where the corresponding bulk surface is no longer the same as in Einstein gravity \cite{BuenoMyers}. This would modify the 
functional form of $\q(\theta)$ and hence the lower bound might be violated for some values of the bulk couplings. 

A more speculative suggestion comes from the observation that holographic models describe CFTs at infinite coupling. It is tempting to guess that for general CFTs, the corner function should lie between those of the holographic models and the free scalar field in  Fig.\ts\ref{fig1}. While the Wilson-Fisher CFTs are outliers at present, this possibility can still be accommodated by the estimated error bars in Table \ref{tbl:ratio}. Improved numerical results might soon clarify this situation \cite{RMelko}.

\textit{Acknowledgments:} We are thankful to H.~Casini for providing the free field results used in Fig.\ts\ref{fig1},  
as well as the integrals for $\sigma_{\rm scalar}$ and $\sigma_{\rm fermion}$. We are grateful for many useful exchanges 
with J.~Cardy, H.~Casini, E.~Fradkin, T.~Grover, R.~Melko, M. Smolkin, M.~Stoudenmire and M.~Taylor. PB thanks Perimeter's Visiting Graduate Fellows Program 
and Perimeter Institute, where part of the project was carried out.
Research at Perimeter Institute is supported by the Government of Canada through Industry Canada and by the Province of Ontario through the Ministry of Research \& Innovation. The work of PB has been supported by the JAEPre grant 2011 00452. RCM acknowledges support from an NSERC Discovery grant and funding from the Canadian Institute for Advanced Research.

\onecolumngrid  \vspace{1cm} 
\begin{center}  
{\Large\bf Supplementary Information} 
\end{center} 
\appendix 
\tableofcontents

\section{Holographic calculations} \labell{holo}

In holographic theories dual to Einstein gravity, the EE is computed using the Ryu-Takayanagi 
prescription  \cite{Ryu:2006bv,Ryu:2006ef}. This states that the EE of a region $V$ in the $d$-dimensional boundary of an asymptotically 
AdS$_{d+1}$ spacetime is proportional to the \emph{area} $\mathcal{A}(\gamma)$ of the codimension-2 bulk surface $\gamma$ which is homologous to $V$ in the boundary (in particular, $\partial \gamma=\partial V$) and extremizes the area functional. That is, we have  
\beq
\see(V)=\underset{\gamma\sim V}{\text{ext}}\left[\frac{\mathcal{A}(\gamma)}{4G} \right]\, .\labell{RyuTaka}
\eeq
We are interested in the vacuum state of various three-dimensional holographic theories, so our bulk geometry will be four-dimensional Euclidean anti-de Sitter space, which we write in Poincar\'e coordinates as 
\begin{align}
  ds^2=\frac{\tilde{L}^2}{z^2}(dz^2+dt_{\ssc E}^2+d\rho^2+\rho^2 d\phi^2)\,, 
\end{align}
where $t_{\ssc E}$ is the Euclidean time. 
This metric is a solution of Einstein gravity coupled to a negative cosmological constant, 
\beq
I_0=\frac{1}{16\pi G}\int d^4x \sqrt{g} \left[\frac6{L^2}+R \right]\, ,
\eeq
 provided we identify both length scales, $\tilde{L}=L$. We can compute the HEE for a region $V$ delimited by two straight lines which intersect at the origin forming a corner with opening angle $\theta$: $V= \left\{\te=0,\,\rho>0,\,|\phi|\le  \theta/2  \right\}$. The bulk surfaces $\gamma$ can be parameterized as $z=\rho\, h(\phi)$, where $h(\phi)$ approaches $\delta/\rho$ as $\phi\rightarrow \pm \theta/2$, with $\delta$ being the UV cut-off. By symmetry, we also have $h(\phi)=h(-\phi)$ and hence $\partial_{\phi}h|_{\phi=0}=0$. The final result for the HEE, obtained using \req{RyuTaka}, reads
\beq 
\see=\frac{\tilde{L}^2}{2G}\,\frac{\ell}{\delta}-a_{\ssc E}(\theta)\,\ln\!\left({\ell}/{\delta} \right) +\mathcal{O}(1) \, ,\labell{eek6}
\eeq
where we have introduced an IR regulator scale, $\rho_{\rm\ssc max}=\ell$, to ensure that the EE does not diverge. 
The function $\qe(\theta)$ is implicitly given by the following expressions \cite{Hirata:2006jx}
\beqa
a_{\ssc E}(h_0)&=&\frac{\tilde{L}^2}{2G}\int_{0}^ {\infty} dy \left[1-\sqrt{\frac{1+h_0^2(1+y^2)}{2+h_0^2(1+y^2)}}\, \right]\, , \labell{q3}\\
 \labell{om} 
\theta(h_0)&=&\int_{0}^{h_0}dh \,\frac{\,2h^2\sqrt{1+h_0^2}}{\sqrt{1+h^2}\sqrt{(h_0^2-h^2)(h_0^2+(1+h_0^2)h^2)}}\,.
\eeqa

When bulk gravity theory contains higher curvature terms, the Ryu-Takayanagi prescription for HEE must be revised. In particular, the area functional in \req{RyuTaka} must be replaced by a new gravitational entropy functional (see \eg \cite{highc2,Dong,Camps:2013zua}), just like the Bekenstein-Hawking formula for the black-hole entropy is replaced by Wald entropy \cite{Wald:1993nt,ted9,wald2} in that case. Schematically, we have 
\beq
\see(V)=\underset{\gamma\sim V}{\text{ext}}S_\mt{grav}(\gamma)\, ,
\labell{newer} 
\eeq
where the new functional depends on the details of the higher curvature action. The form of $S_\mt{grav}$ for the action considered in the main text
\begin{eqnarray}\labell{fo}
I= \int \frac{d^{4}x \, \sqrt{g}}{16\pi G}\left[\frac{6}{L^2}+R+ L^4\lambda_1 R \mathcal{X}_4+L^6\lambda_2 \mathcal{X}_4^2\right]\,,
\end{eqnarray}
is given by \cite{Sarkar,BuenoMyers} 
\beq \labell{see3x} 
S_\mt{grav} \!=\!
\int_{\gamma }\!\! \frac{d^2y\, \sqrt{h}}{4G}\Big[1+\lambda_1 L^4 (\mathcal{X}_4+2R\mathcal{R})+4\lambda_2L^6\mathcal{X}_4\mathcal{R}\Big] \, .
\eeq
When evaluated on empty AdS$_4$, we find $R=-12/\tilde{L}^2$, $\mathcal{X}_4=24/\tilde{L}^4$. Further, it is not difficult to show that the combination $\sqrt{h}\mathcal{R}$ is a total derivative. In particular, one finds
\beq
\sqrt{h}\mathcal{R}=\frac{2\left(-(1+2h^2)\partial_{\phi}h^2- \partial_{\phi}h^4+(h+h^3)\partial^2_{\phi}h\right)}{\rho h^2\left(1+h^2+\partial_{\phi}h^2 \right)^{3/2}}=\frac{d}{d\phi}\left[ \frac{2}{\rho}\frac{\partial_{\phi}h}{\sqrt{1+h^2+\partial_{\phi}h^2}}\right]\, .
\eeq
Using these results, one finds that the extremal bulk surface is unchanged and that $a(\theta)$ is only modified by an overall factor, 
\beq
a(\theta)=\alpha\, a_{\ssc E}(\theta)\qquad{\rm with}\ \ \alpha=1+24\lambda_1+O(\lambda_i^2)\,.
\eeq

Let us close this section by mentioning that we have actually computed $a(\theta)$ and $\ctt$ for the following broader class of higher-curvature theories -- see \cite{BuenoMyers} for details,
\begin{eqnarray}\labell{foo}
&&I=\frac{1}{16\pi G}\int d^{4}x \, \sqrt{g}\left[\frac{6}{L^2}+R+L^2\left(\lambda_1 R^2+\lambda_2 R_{\mu\nu}R^{\mu\nu}+\lambda_{\text{GB}} \mathcal{X}_4\right)\right.\\ \notag 
&&\qquad\qquad+L^4 \left(\lambda_{3,0}R^3+\lambda_{1,1}R\mathcal{X}_4\right) +L^6 \left(\lambda_{4,0}R^4+\lambda_{2,1}R^2\mathcal{X}_4+\lambda_{0,2}\mathcal{X}_4^2\right) \bigg] \,\, .
\end{eqnarray}
Note that the couplings $\lambda_{1,1}$ and $\lambda_{0,2}$ above correspond to $\lambda_1$ and $\lambda_2$ in \req{fo}, respectively.
The final expression of the corner coefficient and the corresponding charges $\sigma$ and $\ctt$ take the form
\beq
a(\theta)=\alpha\,a_{\ssc E}(\theta)\, , \quad \sigma=\alpha\,\sigma_{\ssc E}\, ,\quad{\rm and}\quad  \ctt=\alpha\,\ctte\, ,
\labell{finfin}
\eeq
where to leading order in the dimensionless couplings, the overall coefficient is given by
\beq
\labell{chacha}
\alpha=1-24\lambda_1-6\lambda_2+432\lambda_{3,0}+24\lambda_{1,1}-6912\lambda_{4,0}-576\lambda_{2,1}+\mathcal{O}(\lambda^2)\, .
\eeq
Hence, we have
\beq
\frac{\q(\theta)}{\ctt}=\frac{\qe(\theta)}{\ctte}\qquad{\rm and}\qquad
\frac{\sigma}{\ctt}=\frac{\sge}{\ctte}=\frac{\pi^2}{24}\, 
\eeq
for all the boundary CFTs which are dual to \req{foo}.

\section{Field theory calculations of $\sigma$} \labell{cheese2}

 The first fourteen coefficients in the Taylor expansion of $a(\theta)$ around $\theta=\pi$ were computed numerically for the cases of a free massless scalar and a free massless Dirac fermion using quantum field theory techniques in \cite{Casini:2009sr,Casini:2006hu,Casini:2008as}. The first nonvanishing coefficients correspond to $\sigma_{\rm scalar}$ and $\sigma_{\rm fermion}$, which can be obtained by evaluating the following 
complicated integrals

\begin{eqnarray}\label{int1}
\sigma_{\rm scalar}&=&-2 \pi\,\int_{1/2}^{+\infty} dm \int_{0}^{+\infty}db\  \mu\, H\,a (1-a) \,m\,  \text{sech}^2(\pi  b)\, ,\\ \label{int2}
\sigma_{\rm fermion}&=&-4\pi\int_{1/2}^{+\infty} dm \int_{0}^{+\infty}db\, \left[\mu \,H\, a(1-a)-\frac{F}{4\pi}\right]\,m \, \text{cosech}^2(\pi b)\, ,
\end{eqnarray}
where 
\begin{eqnarray}
\notag
H&\equiv&-\frac{c}{2h} X_1T-\frac{1}{2c} X_2T+\frac{1}{16 \pi  a (a-1) }\, , \\ \notag
h&\equiv&\frac{2 \left(a(a-1) +m^2\right) \sin ^2(\pi  a)}{m^2 \left(\cos (2 \pi  a)+\cos \left(\pi  \sqrt{1-4 m^2}\right)\right)}\, ,\\ \notag
c&\equiv&\frac{ 2^{2 a-1}\pi a (1-a) \, \sec \left(\frac{\pi}{2} \left(2 a+\sqrt{1-4 m^2}\right)\right)\,\Gamma\! \left(\frac{3}{2}-a+\frac12\sqrt{1-4
   m^2}\right)}{m\, \Gamma (2-a)^2\, \Gamma\! \left(a-\frac12+\frac{1}{2} \sqrt{1-4 m^2}\right)}\, ,\\ \notag
X_1&\equiv&-\frac{ \Gamma (-a) \left[\pi  \sinh \left(\frac{\pi  \mu }{2}\right)+ i \cosh \left(\frac{\pi  \mu }{2}\right) \left(\psi^{(0)}\!\left(a+\frac{i \mu }{2}+\frac{1}{2}\right)-\psi ^{(0)}\!\left(a-\frac{i \mu }{2}+\frac{1}{2}\right)\right)\right]}{2^{2 a+1}\mu \, \Gamma (a+1)\, \Gamma\!
   \left(-a-\frac{i \mu }{2}+\frac{1}{2}\right)\, \Gamma\! \left(-a+\frac{i \mu }{2}+\frac{1}{2}\right) (\cos (2 \pi  a)+\cosh (\pi  \mu ))}\,,\\ \label{ess}
   X_2&\equiv&{\rm ``}X_1{\rm "}\ \,\, {\rm with } \,\,\, a\,\, \, {\rm replaced\,\,\, by }\, \,\, (1-a),\, \\ \notag
T&\equiv& \sqrt{h(a^2-a+(h+1)m^2)}\, ,\\ \notag
F&\equiv&-\frac{F_1}{F_2}\, ,\\ \notag
F_1&\equiv & a^2 \left(8 \pi  c^2 \left(m^2+1\right) X_1 T+8 \pi  h
   \left(m^2+1\right)X_2T-c h\right)-16 \pi  a^3T \left(c^2
   X_1+h X_2\right)\\ \notag&+&a \left(-8 \pi  c^2 m^2 X_1T-8 \pi h m^2 X_2T+c h\right)+8 \pi  a^4 T \left(c^2 X_1+h X_2\right)-c h (h+1) m^2\, ,\\ \notag
F_2&\equiv &\frac{8 c\, h \left(a^2-a+m^2\right)^2}{(2 a-1) \mu}\, ,\\ \notag
\mu&\equiv& \sqrt{4m^2-1}\, ,\\ \notag
%
a&\equiv&\left\lbrace \begin{array}{cll} 
 i\,b  +\frac{1}{2}&&\text{for the scalar}\, ,\\ 
i\,b && \text{for the fermion}\, , \end{array}\right.
\end{eqnarray}
and $\psi ^{(0)}$ denotes the digamma function  \ie $\psi ^{(0)}(z)=\frac{d\ }{dz}\ln\Gamma(z)$.
Notice that Eqs.~(\ref{int1}) and (\ref{int2}) look very different and without further insight, we find no reason to believe that these integrals should produce either simple or similar results.

It is possible to compute integrals (\ref{int1}) and (\ref{int2}) numerically with arbitrary precision (although, of course, the computation time increases considerably as we increase the precision). Our results indicate that both Eqs.~(\ref{int1}) and (\ref{int2}) exactly produce the results predicted assuming that $\sigma/\ctt$ is given by the universal constant $\sigma/\ctt=\pi^2/24$, \ie
\beq
\sigma_{\rm scalar}=\frac{1}{256}=0.00390625\,,\qquad
\sigma_{\rm fermion}=\frac{1}{128}=0.0078125\,.
\eeq
We  have verified this result  to a precision of approximately one part in $10^{12}$. In particular, we find
\beq
\sigma_{\rm scalar}=0.00390625000000(5)\,,\qquad
\sigma_{\rm fermion}=0.00781250000000(7)\,,
\eeq
where the numbers in brackets are out of the range of accuracy of our computation.
Let us note that the fact that the previous numerical results seemed to satisfy $\sigma_{\rm fermion}\!=\!2\sigma_{\rm scalar}$ was observed in \cite{Casini:2008as,Casini:2006hu,Casini:2009sr}, but no explanation was given. According to our conjecture, the reason comes simply from the well-known result that $C_{\ssc T}^{\rm fermion}=2C_{\ssc T}^{\rm scalar}$ 
 \cite{Osborn:1993cr}.

In fact,  R\'enyi entropies contain regulator independent $a_n(\theta)$ functions analogous to $a(\theta)$, which corresponds to $n=1$. In the regime of a nearly smooth entangling surface, these R\'enyi corner functions also behave as
\beq
a_n(\theta\to \pi )=\sigma_n\  (\pi-\theta)^2\, ,
\eeq
which straightforwardly generalizes the definition of $\sigma$ to $n \neq 1$. 
The general expression we obtain for $\sigma_n$, corresponding to the $n$th R\'enyi entropy (with $n>1$) is given by \cite{Casini:2006hu}
\begin{equation}
\sigma_n=-\sum_{k=1}^{n-1}\frac{4k(n-k)}{n^2(n-1)}\int_{1/2}^{\infty}dm\, m\sqrt{m^2-1/4}\, H_{k/n}\, ,
\end{equation}
where $H_{k/n}$ takes the same form as $H$ in \req{ess} but with $a=k/n$ instead. For $n=2$ and 3, we find
\beqa
\sigma_2&=&-\int_{1/2}^{\infty}dm\, m\sqrt{m^2-1/4}\,H_{1/2}=0.002110857992548703571747488816869325810507(4)\, ,\nonumber\\
\sigma_3&=&-\frac{4}{9}\int_{1/2}^{\infty}dm\, m\sqrt{m^2-1/4}\,\left(H_{1/3}+H_{2/3}\right)= 0.001701632393277135955281871908373362473448(7)\, .\labell{pop}
\eeqa
These numerical results seem to fit the analytic following expressions 
\begin{equation}
\sigma_2=\frac{1}{48\pi^2}\, ,\qquad \sigma_3=\frac{1}{108\pi\sqrt{3}}\,,
\end{equation}
within the numerical accuracy range -- these analytic expressions were originally inferred with 20 significant digits in \req{pop} and then tested by extending the accuracy to 40 significant digits. With these expressions, the corresponding ratios are given by
\begin{equation}
\frac{\sigma_2}{\ctt}=\frac{2}{9}\,,
\qquad \frac{\sigma_3}{\ctt}=\frac{8\pi}{81\sqrt{3}}
\, .\labell{sigg2}
\end{equation}
As noted in the main text, computing these quantities, and more generally $\sigma_n/\ctt$, for other theories and investigating whether the results in \req{sigg2} are universal, would be of great interest.

\section{Corner coefficient $\q(\theta)$ for the extensive mutual information model} \labell{cheese3}

In this section, we outline the derivation of the corner coefficient $\q_{\rm \ssc Ext}(\theta)$, Eq.\ts\req{ansatz2}, in the  
so-called \emph{extensive mutual information} model of \cite{Casini:2005rm,Casini:2008wt}. 
As its name suggests, this model is characterized by the special property that the mutual information $I(A,B)$ satisfies the extensivity property: $I(A,B\cup C)=I(A,B)+I(A,C)$.
We note that the expression for $\q_{\rm \ssc Ext}(\theta)$ 
first appeared in Ref.~\cite{Casini:2008wt}, and was later independently obtained by Swingle \cite{Swingle:2010jz}.
Here, we shall follow Swingle's heuristic presentation in terms of twist operators, but emphasize that the end result is the same.
While this model can be extended to any number of spacetime dimensions, we only consider $d=3$.
A key role is played by the twist operator $K_n[V]$, which is a line-operator that introduces a branch cut at the boundary $\partial V$ in the $n$-fold replicated theory. The expectation value of $K_n$ in the $n$-replicated theory yields  
$\langle K_n[V]\rangle={\rm Tr}(\rho_V^n)$, as required to calculate the $n$th R\'enyi entropy.  Inspired by results for twist operators in $d=2$ CFTs, 
Swingle made the following ansatz for the twist field in higher dimensions:
\begin{align} \label{Kn}
  K_n[V] = \exp\left( i\alpha_n \int_{\partial V} \hat n\cdot \vec\phi\right)\, ,
\end{align}
where $\hat n$ is the (spatial) unit vector normal to the boundary, and $\vec\phi$ is a two-component vector field (in $d=3$) defined on $\partial V$. For simplicity, $\vec \phi$ is assumed to be a Gaussian field, and thus entirely determined by its two-point function. The latter can be taken to be
\begin{align} \label{phi2pt}
  \ang{\,\phi^i(\vec x)\,\phi^j(0)\,} = \frac{b_1\,\de^{ij} + b_2\, \hat x^i \hat x^j}{|\vec x|^{2}}\, ,
\end{align}
where $b_i$ are real coefficients and $\hat x^i$ denotes the components of the unit vectors $\hat x = \vec x/|\vec x|$. In contrast to \cite{Swingle:2010jz}, we shall keep track of how the coefficients, $b_1$ and $b_2$, enter in the final answer for the EE and we have also extended the correlator \reef{phi2pt} to include the $b_2$ term. The EE is given by $S(V)=-\partial_n\langle K_n[V]\rangle|_{n=1}$, which simplifies to the following double integral over $\partial V$ 
\begin{align}
  S(V)=\frac{1}{2}\pd_n(\al_n^2)\big|_{n=1} \,\int_{\pd V}\int_{\pd V} \hat n_i^{(1)} \hat n_j^{(2)} \ang{\phi^i(\vec x_1)\,\phi^j(\vec x_2)}\,,
\end{align}
by virtue of \req{Kn} and the Gaussianity assumption.

Specializing to the case of a wedge shaped region with opening angle $\theta$, the integral can be explicitly evaluated and yields a result of the expected form 
\begin{align}
  S(V)=B\frac{\ell}{\de} -\q(\theta)\ln\left(\ell/\de\right) + O(\de/\ell)^0\, ,
\end{align}
where $\de$ is a UV cutoff, $\ell$ an IR scale, and where the corner coefficient reads 
\begin{align} \label{swingle-full}
  \q(\theta) &= \frac{2b_1+b_2}{2}\ \partial_n(\alpha_n^2)\big|_{n=1} \  (1+ (\pi-\theta) \cot\theta) \,,
\end{align}
which holds for $0\leq \theta < 2\pi$. 
Hence both the $b_1$ and $b_2$ terms in the correlator \reef{phi2pt} make the same contribution to the logarithmic term up to an overall factor.  

Following \cite{Swingle:2010jz}, we interpret $\alpha_n^2$ as the scaling dimension of the twist operator and hence we can apply the recent result of  \cite{Hung:2014npa}  
\begin{align}
  \partial_n(\alpha_n^2)\big|_{n=1} = \frac{\pi^3}{24}\ctt\, ,
\end{align}
which holds for general CFTs in $d=3$ -- see also \cite{Perlmutter:2013gua}. Substituting this expression into \req{swingle-full}, we set 
\begin{align}
  2b_1+b_2=6/\pi
\end{align}
to produce the correct normalization for $\theta$ near $\pi$, \ie $\q(\theta)\simeq \pi^2\ctt\,(\pi-\theta)^2/24$. Hence we find
\begin{align} \label{ansatz2}
  \q_{\rm \ssc Ext}(\theta)= \frac{\pi^2}{8} \ctt \left(1 + (\pi-\theta)\cot\theta \right)\, ,
\end{align}
which is plotted in Fig.2 of the main text. We also note that, as previously observed in \cite{Casini:2008wt,Swingle:2010jz}, the above corner coefficient diverges as $\sim \kappa/\theta$ for $\theta\to 0^+$. The coefficient can be easily determined:
\begin{align}
  \kappa = \frac{\pi^3}{8}\ctt\,,
\end{align}
which yields a distinct ratio for $\kappa/\ctt$ compared to the free scalar and Dirac fermion field theories, as well as to the holographic CFTs. 
Further, it can be checked that $\q(\theta)$ satisfies the non-trivial inequalities:
\begin{align}
\q(\theta)\ge0\,,\qquad\q'(\theta)\le0\,,\qquad  \q'(\theta) +\sin(\theta)\q''(\theta)\geq 0\,,
\end{align}
obtained using strong subadditivity of entanglement and Lorentz invariance \cite{Casini:2008as}.  

It is not clear at present why the simple extensive mutual information model, when interpreted in terms of twist operators, 
captures the correct factor of $\ctt$ in the $\theta\to\pi$ limit.  
One non-universal property of this model is that $\ctt$ will appear as an overall pre-factor of the EE associated with \emph{any} region $V$. This includes the case where $V$ is a disk, which in turn implies that $\ctt$ fixes the RG monotone $F$.
In this case, the EE is easily obtained as:
\begin{align} \label{disk_swingle}
  S(\text{disk}) &= \mc B\, \frac{R}{\delta} - F_{\rm \ssc Ext} + O(\delta/R) \\
  F_{\rm \ssc Ext} &= \frac{\pi^2}{2}\,(2b_1+b_2) \, 
  \partial_n(\alpha_n^2)\big|_{n=1}\,,
\end{align} 
where $R$ is the radius of the disk. 
Ref.\ts\cite{Swingle:2010jz} found the same expansion for $R\gg\delta$,
but did not specify the form of the regulator independent constant $F$. Here we not only identify its sign, but also its value in terms of the model parameters. In particular, 
we see that $F_{\rm \ssc Ext}$ contains the same combination $(2b_1+b_2)$ that appeared above in $\q_{\rm \ssc Ext}(\theta)$. Hence using 
the same values for that linear combination and the $n$-derivative, we obtain: 
\begin{align}
  F_{\rm \ssc Ext}= \frac{\pi^4}{8}\, \ctt\,,\labell{house3}
\end{align}
which is consistent with the fact that $F>0$ for a CFT. We note that for a CFT holographically dual to pure Einstein gravity,
$F_{\ssc E}=\frac{\pi^3}{6}\ctte$. Of course, the latter differs from \req{house3} but as illustrated in the main text, there is no universal relation between $F$ and $\ctt$. Nevertheless, CFTs described by the extensive mutual information model 
\footnote{The only known example of this is that of massless 2d free fermions \cite{Casini:2005rm}.} and those dual to Einstein gravity seem to share the non-generic feature that $\ctt$ controls the EE for any region and so it would be interesting to see whether
a deeper connection exists between this model and holography.

\end{document}